\documentclass{article}
\usepackage[letterpaper, portrait, margin=1in]{geometry}

\usepackage{markdown}
\usepackage{authblk}

\usepackage[skins, breakable]{tcolorbox}
\usepackage{xcolor}
\definecolor{teal}{HTML}{008080}
\usepackage{graphicx}
\usepackage{amssymb}
\usepackage{amsmath}
\usepackage{amsthm}
\usepackage{braket}
\usepackage{mathabx} 
\usepackage{hyperref}
\hypersetup{
   colorlinks=true,
   citecolor=teal   
}

\newtcolorbox{mybox}
{
  enhanced,
  rounded corners, 
  width = \textwidth,
  overlay={\node at (frame.south east) {\includegraphics[scale=0.1]{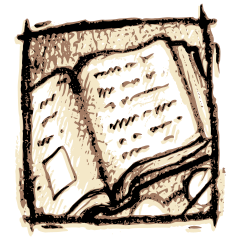}};}
}

\title{The Qudit Cirq Library: An Extension of Google's Cirq Library for Qudits}

\author{Svyatoslav Kushnarev}
\author{Hassan Jameel Asghar}
\affil{\texttt{kusnarevsvatoslav@gmail.com}\\ \texttt{hassan.asghar@mq.edu.au} \\ Macquarie University, Australia}

\date{\today}

\begin{document}

\maketitle

\begin{abstract}
This document contains a guide on how to use the Qudit Cirq library, an extension of Google's Cirq library for qudits. While Cirq provides necessary building blocks for quantum computation on qudits, the gates, curcuits and simulation of qudit systems need to be developed from scratch. Our extension to the Cirq library has built-in methods for creating qudits, applying common qudit gates, building circuits and simulating quantum computation on these circuits. This is ideal for researchers who want to use an off-the-shelf library to simulate qudit systems. 
\end{abstract}

\section{Introduction}
This tutorial provides a guide on how to use the Qudit Cirq Library (\url{https://qudite.pages.dev/}), an extension of Google's Cirq\footnote{See \url{https://quantumai.google/cirq}.} library for qudits, including creating qudits, applying qudit gates, building circuits, and simulating quantum computations with qudits. Qudits are $d$-level quantum systems which serve as the generalization of qubit systems ($d =2$). Theoretically, we assume that they lie in the $d$-dimensional complex Hilbert space $\mathbb{C}^d$. Assume that we have the computational basis $\{\ket{s} : s \in \mathbb{Z}_d\}$ of $\mathbb{C}^d$, where $\ket{s}$ is a column vector of $d$ dimensions with zeros everywhere except at position $s$, where it is $1$. The general form of a qudit $\ket{\psi}$ is:

$$
\ket{\psi} = \sum_{j = 0}^{d-1} \alpha_j \ket{j},
$$

with the condition that $\sum_{j = 0}^{d - 1} |\alpha_j|^2 = 1$, and $\alpha_j \in \mathbb{C}$. For an introduction to qudit systems, see for example~\cite{asghar2024quantum}, and the reference there in. During the development of this library, a couple of other libraries for qudit systems have also been released. One is the QuForge library, which is a standalone library for simulating qudit systems~\cite{quforge}, and the other is the MQT Qudits toolkit for mixed-dimensional quantum
systems~\cite{mqt-qudits} as part of the Munich Quantum Toolkit (MQT)~\cite{mqt} collection of software and tools for quantum computing. Our library builds on top of Google's Cirq, and should be straightforward to use for researchers already familiar with Cirq. 

\section{Creating Qudits}

Qudits are created using Cirq's \verb+LineQid+ class with a specified dimension. The following code creates the qudit $\ket{0}$ of dimension $d = 10$

\begin{mybox}
\begin{verbatim}
import cirq

# Create a qudit of dimension d = 10
qudit = cirq.LineQid(0, dimension=10)
\end{verbatim}
\end{mybox}

To create any other initial state $\ket{s}$ for $s \in \mathbb{Z}_d$ you can use the following code:

\begin{mybox}
\begin{verbatim}
import cirq
import numpy as np 

d = 10
s = 3
initial_state = np.zeros(d, dtype=complex)
initial_state[s] = 1
\end{verbatim}
\end{mybox}

\section{Qudit Gates}

The Qudit Cirq Library provides implementation of common quantum gates.

\subsection*{Qudit Pauli-$X$ Gate (\Verb+quditXGate+)}

The qudit Pauli-$X$ gate generalizes the bit-flip operation to
$d$ dimensions. Its operation on the computational basis state $\ket{s}$ is defined as:

$$
X\ket{s} = \ket{s + 1}
$$

\begin{mybox}
\begin{verbatim}
from qudit_cirq.qudit import quditXGate

# Create a qudit X gate for dimension d=3
x_gate = quditXGate(d=3)
\end{verbatim}
\end{mybox}

This can be applied on a qudit using \verb+.on()+.

\subsection*{Qudit Pauli-$Z$ Gate (\Verb+quditZGate+)}

The qudit Pauli-$Z$ gate generalizes the phase-flip operation. Its operation on the computational basis state $\ket{s}$ is defined as:

$$
Z \ket{s} = \omega^s \ket{s}
$$

\begin{mybox}
\begin{verbatim}
from qudit_cirq.qudit import quditZGate

# Create a qudit Z gate for dimension d=3
z_gate = quditZGate(d=3)
\end{verbatim}
\end{mybox}

This can be applied on a qudit using \verb+.on()+.

\subsection*{Qudit Hadamard Gate (\Verb+quditHGate+)}

The qudit Hadamard gate sends computational basis states into equal superpositions in $d$ dimensions. Its operation on the computational basis state $\ket{s}$ is defined as:

$$
H\ket{s} = \frac{1}{\sqrt{d}}\sum_{j = 0}^{d-1} \omega^{js} \ket{j}
$$

See for example Wang et al.~\cite{wang2020qudits}.

\begin{mybox}
\begin{verbatim}
from qudit_cirq.qudit import quditHGate

# Create a qudit Hadamard gate for dimension d=5
h_gate = quditHGate(d=5)
\end{verbatim}
\end{mybox}

This can be applied on a qudit using \verb+.on()+.

\subsection*{Qudit Controlled-NOT Gate (\Verb+quditCNOTGate+)}

The qudit Controlled-NOT gate generalises the CNOT operation. This is a two qudit operation with two inputs, a control qudit and a target qudit. Given two computational basis states its action is defined as:

$$
\ket{r}\ket{s} \rightarrow \ket{r}\ket{r + s} = \ket{r} X^r \ket{s}
$$

\begin{mybox}
\begin{verbatim}
from qudit_cirq.qudit import quditCNOTGate

# Create a qudit CNOT gate for dimension d=4
cnot_gate = quditCNOTGate(d=4)
\end{verbatim}
\end{mybox}

This can be applied on a qudit using \verb+.on()+. In this case two qudits need to be specified, target and control, respectively.

\subsection*{Qudit $S$ Gate (\Verb+quditPhaseGate+)}

The qudit $S$ gate generalizes the phase gate to dimension $d$. Its operation on the computational basis state $\ket{s}$ is defined as:

$$
S\ket{s} = \omega^{s(s+p_d)/2} \ket{s},
$$

where $p_d = 1$ if $d$ is odd, and $0$ otherwise. This is one of the generators of the qudit Clifford group, along with the Pauli-$Z$ gate, the qudit Hadamard gate and the controlled-$Z$ gate. See for example Jafarzadeh et al~\cite{jafarzadeh2020randomized}.

\begin{mybox}
\begin{verbatim}
from qudit_cirq.qudit import quditPhaseGate

# Create a qudit Phase gate for dimension d=5 (prime and odd)
phase_gate = quditPhaseGate(d=5)
\end{verbatim}
\end{mybox}

This can be applied on a qudit using \verb+.on()+.

\subsection*{Qudit Controlled-$Z$ Gate (\Verb+quditCZGate+)}

The qudit Controlled-$Z$ gate generalises the same operation on qubits. This is a two qudit operation with two inputs, a control qudit and a target qudit. Given two computational basis states its action is defined as:

$$
\ket{r}\ket{s} \rightarrow \omega^{rs}\ket{r}\ket{s} = \ket{r}Z^r \ket{s}
$$

This is one of the generators of the qudit Clifford group, alongwith the Pauli-$Z$ gate, the qudit Hadamard gate and the controlled-$Z$ gate. See for example Jafarzadeh et al~\cite{jafarzadeh2020randomized}.

\begin{mybox}
\begin{verbatim}
from qudit_cirq.qudit import quditPhaseGate

# Create a qudit Phase gate for dimension d=5 (prime and odd)
phase_gate = quditCZGate(d=5)
\end{verbatim}
\end{mybox}

This can be applied on a qudit using \verb+.on()+.

\subsection*{Qudit $U_{\pi/8}$ Gate (\Verb+quditU8Gate+)}

The $U_{\pi/8}$ gate generalizes the qubit $\pi/8$ gate to qudits (called the $T$ gate in the classic text from Chuang and Nielsen [2010]). The extension to qudits is defined for a prime dimension $d$. The general form of the gate is defined as:

$$
U = \sum_{j = 0}^{d-1} \omega^{v_j} \ket{j}\bra{j}
$$

The exact values of the $v_j$'s depends on the dimension $d$. Explicit formula to compuate these values for $d = 3$ and $d \geq 5$ (remember $d$ should be a prime) are given in Howard and Vala~\cite{howard2012qudit}. Our implementation of this gate follows their construction.

\begin{mybox}
\begin{verbatim}
from qudit_cirq.qudit import quditU8Gate

# Create a qudit U_{pi/8} gate for dimension d=7
u8_gate = quditU8Gate(d=7)
\end{verbatim}
\end{mybox}

This can be applied on a qudit using \verb+.on()+.

\section{Building Circuits with Qudits}

There are two primary methods for building circuits with qudits:

\subsection*{Method 1: Manual Construction}

Manually create circuits by explicitly defining qudits and appending gates. Gates are appended to the circuit using the \verb+.append()+ method.

\begin{mybox}
\begin{verbatim}
import cirq
from qudit_cirq.qudit import quditXGate, quditHGate, quditCNOTGate

# Create qudits
qudit1 = cirq.LineQid(0, dimension=3)
qudit2 = cirq.LineQid(1, dimension=3)

# Build a circuit
circuit = cirq.Circuit()
circuit.append(quditHGate(d=3).on(qudit1))
circuit.append(quditCNOTGate(d=3).on(qudit1, qudit2))
\end{verbatim}
\end{mybox}

You can view the circuit using the \verb+print+ command.

\begin{mybox}
\begin{verbatim}
print(circuit)
\end{verbatim}
\end{mybox}

This outputs the following:

\begin{mybox}
\begin{verbatim}
0 (d=3): --H(d=3)--C(d=3)--
                     |
1 (d=3): ----------X(d=3)--
\end{verbatim}
\end{mybox}

\subsection*{Method 2: Using the \Verb+create\_circuit+ Function}

Use the \verb+create_circuit+ function for a more concise syntax when building circuits.

\begin{mybox}
\begin{verbatim}
from qudit_cirq.circuit_builder import create_circuit
from qudit_cirq.qudit import quditHGate, quditCNOTGate, qudit_measure

# Build the circuit using create_circuit
circuit, qudits, qudit_order = create_circuit(
    3,  # Set dimension d=3
    (quditHGate, 'q0'),
    (quditCNOTGate, ['q0', 'q1']),
    (qudit_measure, 'q0'),
    (qudit_measure, 'q1')
)

print(circuit)
\end{verbatim}
\end{mybox}

This yields:

\begin{mybox}
\begin{verbatim}
q0 (d=3): --H(d=3)--C(d=3)--M('m_q0')--
                      |
q1 (d=3): ----------X(d=3)--M('m_q1')--
\end{verbatim}
\end{mybox}

\noindent\textbf{Arguments:}

\begin{enumerate}
    \item The function accepts a variable number of arguments: \verb+\*args+.
    \item Dimension is set by passing an integer.
    \item Gates are specified as tuples with the gate type and qudit names.
\end{enumerate}

\noindent\textbf{Returns:}

\begin{enumerate}
    \item circuit: The constructed \verb+cirq.Circuit+.
    \item qudits: A dictionary mapping qudit names to \verb+cirq.Qid+ objects.
    \item qudit\_order: A list of qudits in the order they were added.
\end{enumerate}

\subsection*{Extended Usage of the \Verb+create\_circuit+ Function}

If needed, the dimensions for each qudit can be specified explicitly in the \verb+create_circuit+ function:

\begin{mybox}
\begin{verbatim}
circuit, qudits, qudit_order = create_circuit(
    (3, quditHGate, 'q0'),
    (4, quditHGate, 'q1'),
    (5, quditTGate, 'q3'),
    (6, quditTGate, 'q4')
)
\end{verbatim}
\end{mybox}

\section{Measurement and Simulation}

Once you have created a circuit, you can measure qudits and simulate the circuit.

\subsection*{Measuring Qudits}

For manual circuits use \verb+.append()+ and \verb+cirq.measure()+ to add measurements to specified qudits. For instance, add the following to the manually constructed circuit above:

\begin{mybox}
\begin{verbatim}
circuit.append(cirq.measure(qudit1, key='q1'))
circuit.append(cirq.measure(qudit2, key='q2'))

print(circuit)
\end{verbatim}
\end{mybox}

This yields:

\begin{mybox}
\begin{verbatim}
0 (d=3): --H(d=3)--C(d=3)--M('q1')--
                     |
1 (d=3): ----------X(d=3)--M('q2')--
\end{verbatim}
\end{mybox}

For circuits built with \verb+create_circuit+, measurements are included during construction as shown above.

\subsection*{Simulating the Circuit}

Use Cirq's simulator to run the circuit and obtain measurement results.

\begin{mybox}
\begin{verbatim}
# Simulate the circuit
simulator = cirq.Simulator()
result = simulator.run(circuit, repetitions=10)

# Print the results
print(result)
\end{verbatim}
\end{mybox}

For the circuit above, this yields:

\begin{mybox}
\begin{verbatim}
q1=2200102011
q2=2200102011
\end{verbatim}
\end{mybox}

\noindent\textbf{Note:}

\begin{enumerate}
    \item \verb+repetitions=10+ specifies the number of times the circuit is run.
    \item result contains the measurement outcomes from all runs.
\end{enumerate}

\section{Utility Functions}

The Qudit Cirq Library provides multiple utility functions to assist with formatting outputs, printing state vectors, and computing tensor products.

\subsection*{Formatting Output (\Verb+format\_out+)}

The \verb+format_out+ function formats a NumPy matrix or vector for display, allowing you to specify the output type for better readability.

\hfill \break
\noindent\textbf{Function definition:}

\begin{mybox}
\begin{verbatim}
def format_out(matrix, output_type='float'):
    # Function implementation
\end{verbatim}
\end{mybox}

\noindent\textbf{Parameters:}

\begin{enumerate}
    \item \verb+matrix+: The NumPy array (matrix or vector) to format.
    \item \verb+output_type+ (optional): The type of formatting to apply to the elements. Options are \verb+float+, \verb+int+, or \verb+str+. Default is \verb+float+.
\end{enumerate}

Example:

\begin{mybox}
\begin{verbatim}
import numpy as np
from qudit_cirq.utils import format_out

# Define a matrix with complex elements
matrix = np.array([[1+0j, 0+0j], [0+0j, -1+0j]])

# Format the matrix as integers
formatted_matrix = format_out(matrix, output_type='int')
print(formatted_matrix)
\end{verbatim}
\end{mybox}

Output:

\begin{mybox}
\begin{verbatim}
[['1' '0']
 ['0' '-1']]
\end{verbatim}
\end{mybox}

\subsection*{Printing the State Vector (\Verb+printVector+)}

The \verb+printVector+ function prints the final state vector of a quantum circuit.

\hfill \break
\noindent\textbf{Function definition:}

\begin{mybox}
\begin{verbatim}
def printVector(final_state, dimensions, qudits=None, threshold=1e-6):
    # Function implementation
\end{verbatim}
\end{mybox}

\noindent\textbf{Parameters:}

\begin{enumerate}
    \item \verb+final_state+: The state vector, e.g., \verb+result.final_state_vector+.
    \item \verb+dimensions+: An integer or list of integers representing the dimensions of the qudits.
    \item \verb+qudits+ (optional): A list of qudit objects to label the qudits in the output.
    \item \verb+threshold+ (optional): A float specifying the minimum amplitude magnitude to display. Default is \verb+1e-6+.
\end{enumerate}

Example:

\begin{mybox}
\begin{verbatim}
import cirq
from qudit_cirq.utils import printVector

d = 3
qudit1 = cirq.NamedQid('q0', dimension=d)
qudit2 = cirq.NamedQid('q1', dimension=d)

circuit = cirq.Circuit()
circuit.append(quditHGate(d).on(qudit1))
circuit.append(quditHGate(d).on(qudit2))


simulator = cirq.Simulator()
result = simulator.simulate(circuit)

final_state = result.final_state_vector

# List of qudits in order
qudit_order = [qudit1, qudit2]

# Print the state vector using printVector
printVector(final_state, dimensions=[d, d], qudits=qudit_order)
\end{verbatim}
\end{mybox}

This should produce the state

$$
\frac{1}{3}\sum_{j = 1}^3 \sum_{k = 1}^3 \ket{jk},
$$

as can be seen from the output:

\begin{mybox}
\begin{verbatim}
Final state vector:
|00⟩: (0.3333333432674408+0j)
|01⟩: (0.3333333432674408+0j)
|02⟩: (0.3333333432674408+0j)
|10⟩: (0.3333333432674408+0j)
|11⟩: (0.3333333432674408+0j)
|12⟩: (0.3333333432674408+0j)
|20⟩: (0.3333333432674408+0j)
|21⟩: (0.3333333432674408+0j)
|22⟩: (0.3333333432674408+0j)
\end{verbatim}
\end{mybox}

\subsection*{Tensor Product of Matrices (\Verb+tensor\_product+)}

The \verb+tensor_product+ function computes the tensor product of multiple matrices or vectors.

\hfill \break
\noindent\textbf{Function definition:}

\begin{mybox}
\begin{verbatim}
def tensor_product(*arrays):
    # Function implementation
\end{verbatim}
\end{mybox}

\noindent\textbf{Parameters:}

\begin{enumerate}
    \item \verb+*arrays+: A variable number of NumPy arrays to compute their tensor product.
\end{enumerate}

Example:

\begin{mybox}
\begin{verbatim}
import cirq
import numpy as np
from qudit_cirq.utils import tensor_product

# Define single-qudit states
state0 = np.array([1, 0, 0])  # |0⟩ in dimension 3
state1 = np.array([0, 1, 0])  # |1⟩ in dimension 3

# Compute the tensor product to create a two-qudit state
combined_state = tensor_product(state0, state1)

print(combined_state)
\end{verbatim}
\end{mybox}

Output:

\begin{mybox}
\begin{verbatim}
[0 1 0 0 0 0 0 0 0]
\end{verbatim}
\end{mybox}

\section{Examples}

We provide a couple of end-to-end examples. These examples create the Greenberger–Horne–Zeilinger (GHZ) state for qudits. The GHZ state for an $n$-qudit system in $d$ dimensions is defined as:

$$
\frac{1}{\sqrt{d}} \sum_{j = 0}^{d-1} \ket{j}^{\otimes n}
$$

The examples below construct this GHZ state using the circuit shown in Asghar, Mukherjee and Brennen~\cite{asghar2024quantum}.

\subsection*{Example 1: Qudit GHZ State Preparation (Manual Construction)}

Prepare a GHZ state of $n = 3$ qudits of dimension $d = 3$.

\begin{mybox}
\begin{verbatim}
import cirq
from qudit_cirq.qudit import quditHGate, quditCNOTGate

# Parameters
d = 3  # Qudit dimension
n_qudits = 3  # Number of qudits

# Create qudits
qudits = [cirq.LineQid(i, dimension=d) for i in range(n_qudits)]

# Build the circuit
circuit = cirq.Circuit()
circuit.append(quditHGate(d).on(qudits[0]))
for i in range(1, n_qudits):
    circuit.append(quditCNOTGate(d).on(qudits[i - 1], qudits[i]))

# Measure
circuit.append(cirq.measure(*qudits, key='result'))

# Simulate
simulator = cirq.Simulator()
result = simulator.run(circuit, repetitions=10)
print(result)
\end{verbatim}
\end{mybox}

Output:

\begin{mybox}
\begin{verbatim}
result=2201011100, 2201011100, 2201011100
\end{verbatim}
\end{mybox}

If instead, we want to print the state vector, then we can replace the measurement and simulation code with:

\begin{mybox}
\begin{verbatim}
simulator = cirq.Simulator()
result = simulator.simulate(circuit)
final_state = result.final_state_vector

printVector(final_state, dimensions=[d, d, d])
\end{verbatim}
\end{mybox}

This gives the output:

\begin{mybox}
\begin{verbatim}
Final state vector:
|000⟩: (0.5773502588272095+0j)
|111⟩: (0.5773502588272095+0j)
|222⟩: (0.5773502588272095+0j)
\end{verbatim}
\end{mybox}

as expected. We can also print this circuit:

\begin{mybox}
\begin{verbatim}
0 (d=3): --H(d=3)--C(d=3)----------
                     |
1 (d=3): ----------X(d=3)--C(d=3)--
                             |
2 (d=3): ------------------X(d=3)--
\end{verbatim}
\end{mybox}

\subsection*{Example 2: Qudit GHZ State Preparation (Using the \Verb+create\_circuit+ Function)}

Prepare the same GHZ state using the \verb+create_circuit+ function.

\begin{mybox}
\begin{verbatim}
import cirq
from qudit_cirq.qudit import quditHGate, quditCNOTGate, qudit_measure

# Build the circuit using create_circuit
circuit, qudits, qudit_order = create_circuit(
    3,  # Set dimension d=3
    (quditHGate, 'q0'),
    (quditCNOTGate, ['q0', 'q1']),
    (quditCNOTGate, ['q1', 'q2']),
    (qudit_measure, 'q0'),
    (qudit_measure, 'q1'),
    (qudit_measure, 'q2')
)

# Simulate
simulator = cirq.Simulator()
result = simulator.run(circuit, repetitions=10)
print(result)
\end{verbatim}
\end{mybox}

Output:

\begin{mybox}
\begin{verbatim}
m_q0=2002211000
m_q1=2002211000
m_q2=2002211000
\end{verbatim}
\end{mybox}

\section{Computational Constraints}

In order to assess the dimension $d$ and the number of qudits $n$ that can be processed in reasonable time via our library, we ran a simple test on a computer with the specification: AMD Ryzen 5 5500U at 2.10 GHz, Radeon Graphics, 8GB RAM. We set a maximum time limit of approximately one minute per simulation run, ensuring that no single circuit execution exceeded this threshold. The circuits were built randomly using the following schema:

\begin{itemize}
    \item Initialise a specified number $n$ of qudits at dimension $d$.
    \item Randomly choose a 10 single-qudit or two-qudit gates from $X$, $Z$, $H$, and $CX$. Thus the circuit depth is 10.
    \item Append measurement operations.
\end{itemize}

The results are shown in the figure below. The reported results are based on single-run scenarios for each dimension and qudit count. Additional runs and averaging could provide more robust metrics.

\begin{center}
\includegraphics[width=0.8\textwidth]{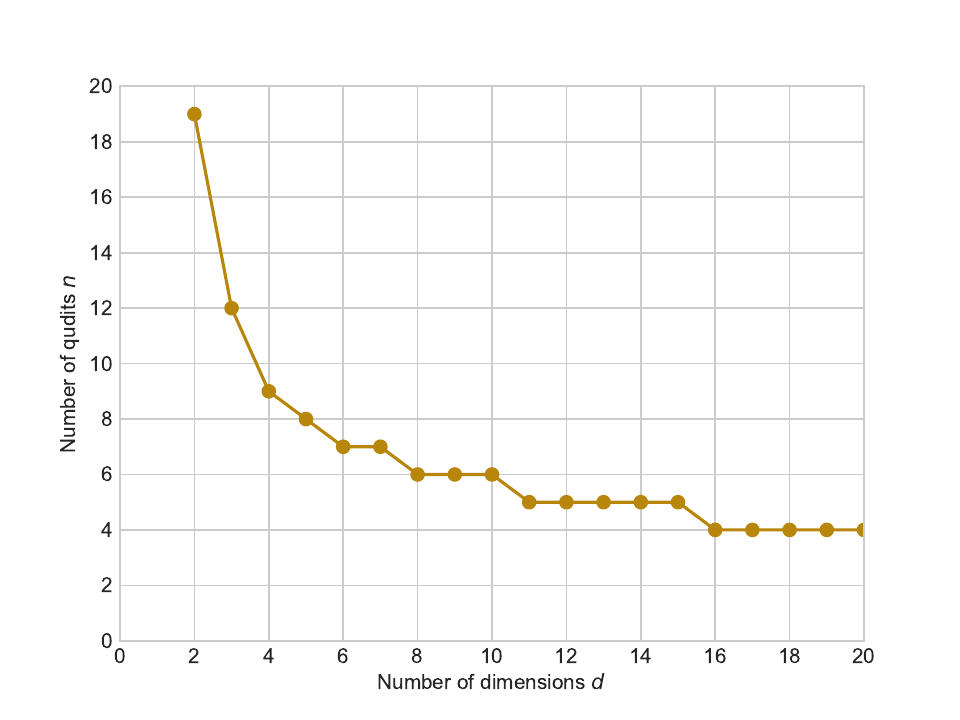}
\end{center}

Not suprisingly, increasing the dimension results in fewer number of qudits processed within the one minute time limit imposed by us. However, by leveraging GPUs and/or quantum cloud computing services the library should be able to handle even larger qudit circuits with greater complexity.

\section*{Acknowledgment}
This work was funded in part by a Future Communications Research Centre grant from Macquarie University.

\bibliographystyle{plain}
\bibliography{qudit-cirq-ref}
\end{document}